\documentclass[conference]{IEEEtran}

\usepackage[noadjust]{cite}
\usepackage{blindtext} 
\usepackage{amsmath} 
\usepackage{amsfonts} 
\usepackage[unit-mode=text, 
]{siunitx}
\usepackage{grffile}
\usepackage{graphicx,epstopdf} 
\usepackage{subfigure} 
\usepackage{tabulary}
\usepackage{xcolor} 
\usepackage[hidelinks]{hyperref}
\usepackage{tikz}

\newcommand{\tsup}{\textsuperscript}
\newcommand{\tsub}{\textsubscript}
\DeclareMathOperator*{\argmax}{arg\,max} 

\newcommand{\rom}[1]{\uppercase\expandafter{\romannumeral #1\relax}}

\newcommand\copyrighttext{%
  \footnotesize \textcopyright 2023 VDE VERLAG GMBH. All rights reserved with the VDE Conference Department.}
  
\newcommand\copyrightnotice{%
\begin{tikzpicture}[remember picture,overlay]
\node[anchor=south,yshift=10pt] at (current page.south) {\fbox{\parbox{\dimexpr\textwidth-\fboxsep-\fboxrule\relax}{\copyrighttext}}};
\end{tikzpicture}%
}    

\begin{document}
	\bstctlcite{BSTcontrol} 
	
	\title{Mobility Performance Analysis of RACH Optimization Based on Decision Tree Supervised Learning for Conditional Handover in 5G Beamformed Networks
 \vspace{-0.7\baselineskip}}
 
	\author{\IEEEauthorblockN{
			Subhyal Bin Iqbal\IEEEauthorrefmark{1}\IEEEauthorrefmark{2},
            Umur Karabulut\IEEEauthorrefmark{1}, 
			Ahmad Awada\IEEEauthorrefmark{1},
			Andre Noll Barreto\IEEEauthorrefmark{3},
            Philipp Schulz\IEEEauthorrefmark{2} and
			Gerhard P. Fettweis\IEEEauthorrefmark{2}}
		\IEEEauthorblockA{
			\IEEEauthorrefmark{1} {Nokia Standardization and Research Lab, Munich, Germany} and
            \IEEEauthorrefmark{3} {Paris, France}\\
			\IEEEauthorrefmark{2} {Vodafone Chair for Mobile Communications Systems, Technische Universit\"at Dresden, Germany}
		}
	\vspace{-1.2\baselineskip}}
 
	\maketitle

%
\copyrightnotice

	\begin{abstract}
		In 5G cellular networks, frequency range 2 (FR2) introduces higher frequencies that cause rapid signal degradation and challenge user mobility. In recent studies, a conditional handover procedure has been adopted as an enhancement to baseline handover to enhance user mobility robustness. In this article, the mobility performance of conditional handover is analyzed for a 5G mm-wave network in FR2 that employs beamforming. In addition, a resource-efficient random access procedure is proposed that increases the probability of contention-free random access during a handover. Moreover, a simple yet effective decision tree-based supervised learning method is proposed to minimize the handover failures that are caused by the beam preparation phase of the random access procedure. Results have shown that a tradeoff exists between contention-free random access and handover failures. It is also seen that the optimum operation point of random access is achievable with the proposed learning algorithm for conditional handover. Moreover, a mobility performance comparison of conditional handover with baseline handover is also carried out. Results have shown that while baseline handover causes fewer handover failures than conditional handover, the total number of mobility failures in the latter is less due to the decoupling of the handover preparation and execution phases.
	\end{abstract}
	

 \vspace{-0.6\baselineskip}
	\section{Introduction}
		In 5G cellular networks, the demand for user data throughput is poised to increase dramatically \cite{Throughput}. To address this, the range of carrier frequencies has been further increased to frequency range 2 (FR2) \cite{38331} to fulfill this ever-increasing demand. However, operating in FR2 challenges user~mobility due to the higher free-space path loss and penetration loss which can lead to rapid signal degradation \cite{HOsurvey}. Moreover, the dense BS deployment in 5G networks increases the total number of handovers, which can cause frequent interruption of the user equipment (UE) connection and increase the signaling~overhead \cite{HOsurvey}. 
		
		Baseline handover (BHO) is a handover procedure used in Long Term Evolution. It has also been reused for 5G networks in the 3\tsup{rd} Generation Partnership Project (3GPP) \textit{Release 15} \cite{38300, 38331}. In BHO, the time instant for triggering the handover is critical because the signal of the serving cell should be good enough to receive the handover command and the signal of the target cell should be sufficient for access. This is more pronounced in FR2 due to the rapid signal degradation and dense BS deployment. 
		
		Conditional handover (CHO) has been introduced in \cite{CHOTechRep} for 5G New Radio \textit{3GPP Release 16} as a handover mechanism to increase the mobility robustness of BHO. In CHO, the coupling between handover preparation and execution is removed by introducing a conditional procedure, whereby the handover is prepared early by the serving cell and access to the target cell is performed only later when its radio link is sufficient. A contention-free random access (CFRA) procedure has already been defined in \cite{38331}. Here the target cell of the handover can allocate CFRA resources for the UE during the handover. Using CFRA instead of contention-based random access (CBRA) resources helps to avoid collision in random access and consequently reduces mobility interruption time and signaling overhead \cite{RAlinktoSingalingOverheadandLatency1, RAlinktoSingalingOverheadandLatency2}. 
 

        The authors in \cite{JedrzejRACHpaper} have proposed a scheme to reduce the number of CFRA attempts during a handover but it focuses on using an additional measurement report between the CHO preparation and execution phases to update the CFRA resources for specific beams. In the first contribution of this article, a resource-efficient random access channel (RE-RACH) procedure is proposed as a simple yet powerful enhancement to the 3GPP RACH model \cite{38331} such that the utilization of CFRA resources is increased. The authors in \cite{MarvinHOFpaper} have also proposed a scheme to reduce HOFs in a mobile environment by using different machine learning techniques including decision tree-based supervised learning but they consider a non-beamformed 5G system with BHO. In the second contribution of this article, a beam-specific enhanced logging and learning (BELL) approach based on a decision tree-based supervised learning algorithm \cite{SupervisedLearning} is proposed. The BELL approach is shown to decrease handover failures (HOFs) by using a learning approach to avoid HOFs caused by wrong beam preparation during the beam preparation phase of the random access procedure. A reduction in HOFs also means unnecessary RACH attempts can be avoided and therefore resource utilization can be improved. To the best of the authors' knowledge, both these enhancements are novel and their mobility performance in 5G cellular beamformed networks has not been investigated in literature before. Later in the article, the RE-RACH and BELL approaches are combined to study the tradeoff between CFRA resource utilization and HOFs and furthermore optimize the mobility performance.


%
	\section{UE Measurements and Handover Models}
	\label{sec:Meas_HO}
	
		In 5G cellular networks it is necessary to hand over the link of a UE between cells to sustain the user connection with the network. This handover is performed using received signal power measurements at the UE from the serving and neighboring cells and by following a predefined handover procedure. In this section, BHO and CHO procedures are reviewed along with the relevant UE measurements for mobility.
				
		\subsection{UE Measurements in 5G Beamformed Networks}
		A UE $u$ in the network monitors the reference signal received power (RSRP) $ P_{c,b}^\textrm{RSRP}(n) $ (in dBm) at discrete time instant $ n $ for beams $b\in B $ of cell $ c \in C$, using the synchronization signal block (SSB) bursts that are transmitted by the base station (BS). $u$ is left out in the subscript here and in future instances for simplicity. The separation between the time instants is given by $ \Delta t $ ms. The raw RSRP measurements are inadequate for handover decisions since those measurements fluctuate rapidly over time due to fast fading and measurement errors which would lead to unstable handover decisions. To mitigate these channel impairments, the UE applies a moving average layer-1 (L1) filter and an infinite impulse response (IIR) layer-3 (L3) filter sequentially to the RSRP measurements. The implementation of L1 filtering is not specified in 3GPP standardization and it is UE specific, i.e., it can be performed either in linear or dB domain. The L1 filter output can be expressed as
  %
		\begin{equation}\label{eq:L1filter}
		P_{c,b}^\textrm{L1}(m) = \frac{1}{N_\textrm{L1}}\sum_{\kappa = 0}^{N_\textrm{L1}-1}P_{c,b}^\textrm{RSRP}(m-\kappa),~ m=n\omega
		\end{equation}
		where $ \omega \in \mathbb{N}$ is the L1 measurement period normalized by time step duration $ \Delta t $ , and $ N_\textrm{L1} $ is the number of samples that are averaged in each L1 measurement period. For cell quality derivation of cell $ c $, the strongest set $ B^\textrm{str}_c $ of beams having measurements above the threshold $ P_\textrm{thr} $ is determined by the UE as
   %
		\begin{subequations}
			\begin{align}\label{eq:BeamConsol1}
				& B^\textrm{str}_c(m) = \{b~|~P_{c,b}^\textrm{L1}(m) > P_\textrm{thr}\}\\
				\intertext{subject to}
				&P_{c,b_i}^\textrm{L1}(m) > P_{c,b_j}^\textrm{L1}(m),~ \forall b_i \in B^\textrm{str}_c,~\forall b_j \in B\setminus B^\textrm{str}_c,\\
				&|B^\textrm{str}_c| \leq N_\textrm{str}, ~N_\textrm{str}\in \mathbb{N^+}.
			\end{align}
		\end{subequations}

  
   %
		The cardinality of the set is denoted by $ \vert\cdot\vert $ and $ N_\textrm{str} $ is the maximum number of beams that are accounted for cell quality derivation. 
        L1 RSRP measurements of beams $ b\in B^\textrm{str}_c $ are averaged to derive the L1 cell quality of cell $ c $ as
   
		\begin{equation}\label{eq:BeamConsol2}
		P_c^\textrm{L1}(m) = \frac{1}{\vert B^\textrm{str}_c(m)\vert}\sum_{b\in B^\textrm{str}_c(m)}P_{c,b}^\textrm{L1}(m). 
		\end{equation}
		If $ B^\textrm{str}_c(m) $ is empty, $ P_c^\textrm{L1}(m) $ is equal to highest $ P_{c,b}^\textrm{L1}(m) $.
		
		L1 cell quality is further smoothed by L3 filtering and L3 cell quality output is derived by the UE as
   %
		\begin{equation}\label{eq:L3Cell}
		P_c^\textrm{L3}(m) = \alpha P_c^\textrm{L1}(m) + (1-\alpha)P_c^\textrm{L3}(m-\omega),
		\end{equation}
		where $\alpha = \left( \frac{1}{2}\right)^\frac{k}{4}$ is the forgetting factor that controls the impact of older measurements $P_c^\textrm{L3}(m-\omega)$ and $ k $ is the filter coefficient of the IIR filter \cite{38331}.
		
		Similarly, the L3 beam measurement $ P_{c,b}^\textrm{L3}(m) $ of each beam is evaluated by L3 filtering of L1 RSRP beam measurements 
   %
		\begin{equation}\label{eq:L3Beam}
		P_{c,b}^\textrm{L3}(m) = \alpha' P_{c,b}^\textrm{L1}(m) + (1-\alpha')P_{c,b}^\textrm{L3}(m-\omega),
		\end{equation}
   
		where $ \alpha' $ can be configured separately from $ \alpha $.
		

				
		\subsection{Baseline Handover}
			
		L3 cell quality measurements $ P_{c}^\textrm{L3}(m) $ are used to assess the quality of the radio links between the UE and its serving and neighboring cells. To this end, the UE reports the L3 cell quality measurements $P_{c}^\textrm{L3}(m) $ and beam measurements $ P_{c,b}^\textrm{L3}(m) $ to its serving cell $ c_0 $ if the following \textit{A3} condition
		\begin{equation}
			\centering
			\label{eq:A3}
			P_{c_0}^\textrm{L3}(m) + o_{c_0,c}^\textrm{A\tsub3}< P_c^\textrm{L3}(m) ~~\textrm{for}~~ m_0-T_\textrm{TTT}^\textrm{A\tsub 3}<m<m_0,
		\end{equation}
		expires at time instant $ m = m_0 $ for any neighboring cell $ c\neq c_0 $. The cell-pair specific handover offset $ o_{c_0,c}^\textrm{A\tsub3} $ can be configured differently by $ c_0 $ for each neighboring cell $ c $ and the time-to-trigger $ T_\textrm{TTT}^\textrm{A\tsub 3} $ is the observation period of condition (\ref{eq:A3}) before a measurement report is initiated by~$ c_0 $. 
		
		After receiving  L3 cell quality measurements from the UE, the serving cell $ c_0 $ sends a \textit{Handover Request} to the target cell~$ c_\textrm{T} $, typically the strongest cell, along with the L3 beam measurements $ P_{c_\textrm{T},b}^\textrm{L3}(m)$. Thereafter, the target cell reserves CFRA resources (preambles) for beams $ b\in B^\textrm{prep}_{c_\textrm{T}} $ with the highest power based on reported $ P_{c,b}^\textrm{L3}(m)$. The target cell~$ c_\textrm{T} $ prepares the \textit{Handover Request Acknowledgement}, which includes the reserved CFRA resources and sends it to the serving cell. Thereafter, the serving cell sends the \textit{Handover Command} to the UE. This command includes the target cell configuration and CFRA preambles that are reserved by the target cell $ c_\textrm{T} $. Upon reception, the UE detaches from the serving cell and initiates random access toward the target cell. 
		
		In this handover scheme, the radio link between UE and the serving cell  should be good enough to send the measurement report in the uplink and receive the handover command in the downlink. In addition, the radio link quality between the UE and the target cell should also be sufficient so that the signaling between UE and the target cell is sustained during the RACH procedure. Herein, the link quality conditions for a successful handover between the serving cell $ c_0 $ and target cell $c_\textrm{T}$ are expressed as
   %
		\begin{subequations}
		\begin{align}
		\gamma_{c_0,b}(m_0) >& \gamma_{\textrm{out}}, \label{eq:ServingCond}\\
		\gamma_{c_0,b}(m_0+T_\textrm{p}) >& \gamma_{\textrm{out}},~~ \label{eq:ServingCond2}\\
		\gamma_{c_\textrm{T},b}(m_0+T_\textrm{p}) >& \gamma_{\textrm{out}}  \label{eq:TargetCond},
		\end{align}
		\end{subequations}
		where $ \gamma_{c,b}(m) $ is the SINR of the link between the UE and the beam $ b $ of cell $ c$. The time instant $ m_0 $ is when the measurement report is sent by the UE and $ T_\textrm{p} $ is the handover preparation time, i.e., the time delay between the UE sending the measurement report and receiving the handover command. $ \gamma_{\textrm{out}} $ is the SINR threshold that is required for maintaining radio communication between UE and the network.

	
		\subsection{Conditional Handover}
		In CHO, the handover preparation and execution phases are decoupled. This helps to receive the handover command safely from the serving cell by triggering the handover early and to access the target cell later when its radio link is sufficient.
		
		Similar to the A3 condition in (\ref{eq:A3}), an \textit{Add} condition is defined as
		\begin{equation}
			\label{eq:Add}
			P_{c_0}^\textrm{L3}(m) + o_{c_0,c}^\textrm{add} < P_c^\textrm{L3}(m)  ~~\textrm{for}~~m_0-T_\textrm{TTT}^\textrm{add}<m<m_0,
		\end{equation}
		where $ o_{c_0,c}^\textrm{add} $ is defined as the \textit{Add} condition offset. The UE sends the measurement report to serving cell $ c_0 $ at $ m = m_0 $ if the \textit{Add} condition is fulfilled for $ T_\textrm{TTT}^\textrm{add} $ seconds. Then, the serving cell $ c_0 $ sends the \textit{Handover Request} to the target cell $ c_\textrm{T} $ for the given UE. The preparation of the handover is performed as in BHO, where the target cell reserves CFRA RACH resources for the UE and sends the \textit{Handover command} to the UE via the serving cell. Unlike BHO, however, the UE does not detach from the serving cell immediately and initiates the RACH process towards the target cell. Instead, the UE continues measuring the RSRP measurements from the neighboring cells and initiates the random access only when the \textit{Execution} condition expires at time instant $ m_1 $
		\begin{equation}
		\label{eq:Exec}
			P_{c_0}^\textrm{L3}(m) + o_{c_0,c_\textrm{T}}^\textrm{exec}< P_{c_\textrm{T}}^\textrm{L3}(m) ~~\textrm{for}~~ m_1-T_\textrm{TTT}^\textrm{exec}<m<m_1.
		\end{equation}
		where $m_1>m_0$. The \textit{Execution} condition offset $ o_{c_0,{c_\textrm{T}}}^\textrm{exec} $ is configured by the serving cell and forwarded to the UE in the handover command along with CFRA resources reserved by the target cell. 
		
		It is observed that larger $ o_{c_0,{c_\textrm{T}}}^\textrm{add} $ values lead to early preparation of the target cell and reservation of the RACH preambles which ensures that the UE can send the measurement report and receive the handover command successfully (cf. (\ref{eq:ServingCond}) and (\ref{eq:ServingCond2})). Besides, unlike BHO, lower $ o_{c_0,{c_\textrm{T}}}^\textrm{add} $ does not lead to any early RACH attempt of the UE towards the target cell since the random access is initiated only when the \textit{Execution} condition is fulfilled. Higher offset $ o_{c_0,{c_\textrm{T}}}^\textrm{exec} $ values cause the UE to perform random access late enough such that it is more likely that the SINR  $ \gamma_{c_\textrm{T},b}(m) $ is above $ \gamma_{\textrm{out}} $ (cf. (\ref{eq:TargetCond})).

	\section{RACH Procedure in 5G Beamfromed Networks}
	\label{sec:RACH}
	In this section, the basics of random access are discussed. Then, the 3GPP RACH procedure used in 5G networks \cite{38331} is described and our proposed RACH procedure is introduced. 
	
	\subsection{Contention-free and Contention-based Random Access}
	
	Random access can be described as the first signaling performed by a UE for establishing the synchronization with a cell during a handover procedure. The UE initiates random access by sending a RACH preamble to the target cell. However, it is possible that multiple UEs use the same preamble during random access towards the same transmit beam of a target cell. In this case, RACH collision occurs, which means that the UE needs to re-transmit the RACH preamble. This results in additional signaling and handover interruption time \cite{RAlinktoSingalingOverheadandLatency1, RAlinktoSingalingOverheadandLatency2}. The type of random access where a UE selects one preamble out of set that is common for all UEs in the network is called~CBRA. 
	
	During a handover, the collision risk can be avoided by assigning dedicated preambles to each UE to be used towards a prepared transmit beam $ b\in B^\textrm{prep}_{c_\textrm{T}} $ of the target cell $ c_\textrm{T} $. The network identifies the UE signal without further signaling and handover interruption time if the UE accesses the prepared beam using the assigned dedicated preamble. This type of random access is called CFRA. Based on \cite{38331}, $ B^\textrm{prep}_{c_\textrm{T}} $ can be defined as the preparation of beams with the strongest RSRP for UE $u$ and can be formulated as
  \vspace{-0.2\baselineskip}
	\begin{subequations}
		\begin{align}\label{eq:PrepBeamSet}
			&B^\textrm{prep}_{c_\textrm{T}}= \{b |P_{c_\textrm{T},b}^\textrm{L3}(m_0) \geq  P_{c_\textrm{T},b_i}^\textrm{L3}(m_0),~b\neq b_i,~b,b_i\in B_{c_\textrm{T}}\}\\
			\intertext{subject to}
			&|B^\textrm{prep}_{c_\textrm{T}}| = N_\textrm{B}
		\end{align}
	\end{subequations}
where $ N_\textrm{B} $ is the number of beams prepared for random access.

	\subsection{Access Beam and Preamble Selection}
	\label{subsec:AccessPreamble}
	During a handover, accessing the target cell by using a dedicated CFRA preamble is preferable due to lower handover interruption time and signaling requirements than for CBRA. Although a set of beams $ b\in B^\textrm{prep}_{c_\textrm{T}} $ of the target cell $ c_\textrm{T} $ with the strongest L3 beam quality measurements $ P_{b,{c_\textrm{T}}}^\textrm{L3} $ can be prepared with CFRA resources, measurements of those beams may vary between the preparation time instant $ m = m_0 $ and access time $ m = m_1  $ due to the decoupling between the phases. The variation of beam measurements is more significant in CHO as compared to BHO. This is because in BHO the elapsed time between the preparation and execution phases is given by the handover preparation time $ T_\textrm{p} $ in (\ref{eq:TargetCond}). However, in CHO there is an additional period of time  $ T_\textrm{f} $ until the \textit{Execution} condition in (\ref{eq:Exec}) is fulfilled after receiving the handover command from the serving cell~$c_0$, which itself follows the CHO preparation given in (\ref{eq:Add}). The time period $ T_\textrm{f} $ depends on the serving and target cell RSRPs and can be as large as \SI{10}{s} \cite{JedrzejRACHpaper}, which therefore makes it much than $ T_\textrm{p}=\SI{50}{ms}$.
	
	\begin{figure}[!t]
	\centering
	\includegraphics[width=0.8\columnwidth]{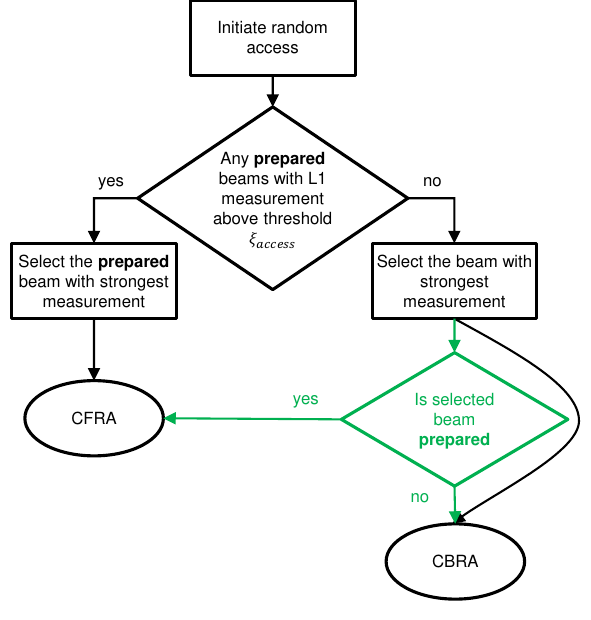}
    \vspace{-16pt}
	\caption{A flow diagram of the random access procedure. The part shown in black is defined in 3GPP standard and the green part is the proposed enhancement for the random access procedure.}
	\label{fig:RACHflow}
	\end{figure}	
	
	Due to this temporal variation of beam measurements, the access beam is selected based on measurements at time instant $ m=m_1 $ of CHO execution. This is illustrated in Fig.\,\ref{fig:RACHflow}. Herein, the UE selects the access beam $ b_{0} \in  B^\textrm{prep}_{c_\textrm{T}} $ as follows
	\begin{equation}\label{eq:SelectBestPrep}
	b_{0} = \argmax_{b  \in B^\textrm{prep}_{c_\textrm{T}} }  P_{{c_\textrm{T}},b}^\textrm{L1}(m_1),~\text{if}~P_{c_\textrm{T},b_{0}}^\textrm{L1}(m_1) > \xi_\textrm{access},
	\end{equation}
	where $ \xi_\textrm{access} $ is the threshold that L1 RSRP beam measurements should exceed to consider prepared beams for access. Ultimately, the UE accesses the prepared beam $ b_0 $ that satisfies the condition in (\ref{eq:SelectBestPrep}) and uses the corresponding CFRA preamble. If none of the measurements $ P_{b,{c_\textrm{T}}}^\textrm{L1} $ of beams $ b\in B^\textrm{prep}_{c_\textrm{T}} $ is above the threshold $ \xi_\textrm{access} $, the UE simply selects any beam $ b_0 $ (as opposed to just the prepared beams in the set $ B^\textrm{prep}_{c_\textrm{T}} $ in (\ref{eq:SelectBestPrep})) with the strongest L1 RSRP beam measurement as
	\begin{equation}\label{eq:SelectStrongBeam}
	b_{0} = \argmax_{b  \in B_{c_\textrm{T}} } P_{{c_\textrm{T}},b}^\textrm{L1}(m_1).
	\end{equation}
		
	In 3GPP standardization \cite{38331}, CBRA preambles are used if none of the L1 RSRP measurement of prepared beams is above the threshold $\xi_\textrm{access}$. This carries the disadvantage that the UE may select CBRA resources although there may be CFRA resources associated with the selected strongest beam. To address this issue, a resource-efficient RACH (RE-RACH) scheme is proposed to increase the CFRA utilization. This enhancement is shown in green in Fig.\,\ref{fig:RACHflow}. Herein, the UE uses CFRA resources if the selected beam happens to be a prepared beam $ b_0\in B^\textrm{prep}_{c_\textrm{T}} $, even if the L1 RSRP beam measurement $ P_{{c_\textrm{T}},b_0}^\textrm{L1} $ is below the threshold $ \xi_\textrm{access} $. This will lead to less signaling and handover interruption during the RACH procedure. The UE selects either CFRA or CBRA preambles by following either of the two RACH procedures shown in Fig.\,\ref{fig:RACHflow} and attempts to access the target cell with the selected preamble. If the random access fails, the UE repeats the preamble selection process and declares a HOF after a predefined number of attempts. This is followed by a re-establishment process in which the UE searches for a new serving cell to be connected to.
	
	\subsection{Beam-specific Enhanced Logging and Learning Approach}

    As mentioned in Section~\ref{subsec:AccessPreamble}, the RSRP of the prepared beams $b\in B^\textrm{prep}_{c_\textrm{T}}$ at time $ m=m_0 $ can change as time elapses due to temporal characteristics of the wireless channel and user mobility. It could be that the RSRP of the prepared beam with the highest RSRP might not be strong enough for successful random access when the access procedure is started at $m=m_1 $ and afterward. As will be seen later in Section~\ref{sec:SimRes}, HOFs are primarily dependent on the radio link quality, and their probability can increase if the quality of the radio link between the UE and the prepared beam is not good enough.
	
\begin{figure}[!b]
	\centering
	\includegraphics[width=0.8\columnwidth]{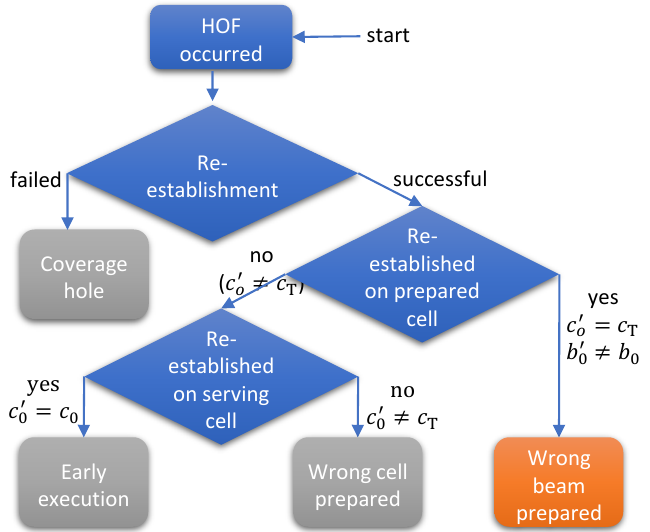}
	\caption{Classification of the HOFs with a decision tree that applies the root-cause analysis on the HOFs. Classes identify the correctness of the decisions on the target cell, prepared beam, and execution time of the handover process along with network planning problems such as coverage holes. The BELL approach focuses on addressing HOFs caused by wrong beam prepared (shown in orange). Serving cell, target cell, re-establishment cell, prepared beam, and re-establishment beam are denoted by $ c_0$, $c_\textrm{T}$, $ c_0' $, $ b_0 $ and $ b_0' $  respectively.} 
	\label{fig:decisiontree}
	\vspace{-0.4\baselineskip}
	\end{figure}

	In this article, we propose a RACH optimization scheme based on a supervised learning algorithm \cite{SupervisedLearning} that aims to learn preparing the correct transmit beam so that the HOFs caused by wrong beam preparation in (\ref{eq:PrepBeamSet}) during the RACH process are minimized. We term it beam-specific enhanced logging and learning  (BELL). BELL classifies the HOF events into sub-events by following a pre-defined decision tree. This is succeeded by an assessment mechanism that reacts to the inferences (classes), either by rewarding or penalizing the network decision on the prepared beam in a way that the decisions leading to successful handovers are encouraged and decisions leading to HOFs are discouraged. The decision tree used in the BELL approach is illustrated in Fig.\,\ref{fig:decisiontree}. It applies the root-cause analysis to classify the HOF events. A handover attempt that is followed by the HOF event could be classified as one of four unique classes, which will be explained next. 

    If the HOF is followed by a re-establishment process that fails, the UE will be unconnected to the network until further connection to a new cell. This type of HOF event is classified as a “coverage hole”. Although the UE failed to hand over to the target cell~$ c_\textrm{T} $, it may connect to cell $ c_0'=c_\textrm{T} $ after the re-establishment process through a beam  $ b_0'\not\in B^\textrm{prep}_{c_\textrm{T}} $ other than the prepared one, which is then interpreted as “wrong beam prepared” (shown in orange). This type of HOF could be avoided if a handover attempt through the wrong beam $ b_0 \in B^\textrm{prep}_{{c_\textrm{T}}} $ is avoided and the preparation of  $ b_0' $ is motivated. The UE may also re-establish on the serving cell $ c_0 $, i.e., $c_0' = c_0$, which shows that the handover execution condition expired before the link between the target cell $ {c_\textrm{T}} $ and UE became good enough for successful handover. Therefore, it is classified as “early execution”. Similarly, if the UE does not connect to target cell $ c_\textrm{T} $ after HOF but rather another cell, i.e., $ c_0'\neq {c_\textrm{T}} $, it shows that the preparation of the target cell $ {c_\textrm{T}} $ was not accurate. This is classified as “wrong cell prepared”. 
	
	It is already mentioned that this article is focused on the RACH optimization by minimizing the HOF caused by the preparation of the wrong beams defined in (\ref{eq:PrepBeamSet}). Therefore, the BELL approach reacts to the HOF events that are classified as “wrong beam prepared”. To this end, a beam-specific preparation offset $ o^\textrm{prep}_{c,b} $ is proposed and equation (\ref{eq:PrepBeamSet}) is reformulated as
  %
	\begin{align}\label{eq:cellbeamoffset}
	B^\textrm{prep}_{c_\textrm{T}} = \{&b |P_{c_\textrm{T},b}^\textrm{L3}(m_0) + o^\textrm{prep}_{c_\textrm{T},b} \geq  P_{ c_\textrm{T},b_i}^\textrm{L3}(m_0)+o^\textrm{prep}_{c_\textrm{T},b_i} , \nonumber \\
	&~b\neq b_i,~b,b_i\in B_{c_\textrm{T}}\}.
	\end{align}
	When a HOF event is classified as “wrong beam prepared”, the beam-specific preparation offset $ o^\textrm{prep}_{c_\textrm{T},b} $ of $ b\in B^\textrm{prep}_{c_\textrm{T}} $ is reduced by $ \Delta o $ dB to penalize the preparation of this beam, and the offset $ o^\textrm{prep}_{c_\textrm{T},b_0'} $ of the re-established beam $ b_0'\not \in B^\textrm{prep}_{c_\textrm{T}} $ is increased by $ \Delta o $ dB to leverage the preparation of the beam $b_0'$ in future events. Enabling such optimization is possible only if the network logs the target cell $ c_\textrm{T} $ and access beam $ b_0 $ information so that the root-cause analysis is applied to classify the HOF events and proper preparation decisions can be learned by the network over~time.
	
	The handover performance can be further optimized and HOF events that are classified as “early execution” and “wrong cell prepared” can be improved.  This can be done by adjusting the cell-specific preparation and execution offsets $ o^\textrm{prep}_{c_0,c} $ and $ o^\textrm{exec}_{c_0,c}  $, respectively, such that the early execution or wrong cell preparation can be discouraged towards a specific cell. However, as will be seen later in Section~\ref{sec:SimRes}, these classes are corner cases that are rarely observed compared to the case “wrong beam prepared”.
 
	
	\section{Simulation Scenario and Parameters}
	\label{sec:SimScenario}
	In this section, the simulation scenario is described along with the simulation parameters which are listed in Table~\ref{tab:sim_parameters}. The simulations have been performed in our proprietary MATLAB-based system-level simulator.


 \begin{table}[!t]
		\renewcommand{\arraystretch}{1.3}
		\caption{Simulation Parameters}
        \vspace{-6pt}
		\label{tab:sim_parameters}
		\centering
		\begin{tabulary}{\columnwidth}{|L | L|}
			
			\hline 
			\textbf{Parameters} & \textbf{Value}\\ 
			\hline \hline
			Carrier frequency & \SI{28}{GHz} \\ 
            \hline
            System bandwidth & \SI{100}{MHz}\\	
		  \hline
			Network topology & Madrid grid \cite{METIS2} \\
			\hline
			Number of cells & $33$\\
            \hline
            Scenario & UMi-Street Canyon \cite{38901}\\
   	    \hline
			PRB bandwidth & \SI{10}{MHz} \\
			\hline
			Downlink Tx power & \SI{12}{dBm/PRB} \\
             \hline   
			Tx antenna height & \SI{10}{m}\\
			\hline
			Tx Antenna element pattern & Table~7.3-1 in \cite{38901} \\
			\hline
			Tx panel size & $16 \times 8,~\forall b\in \{1,\ldots,8\}$ $8 \times 4,~\forall b \in \{9,\ldots,12\}$\\
            \hline
    	   Tx antenna element spacing & vertical: 0.7$\lambda$\\ 
            & horizontal: 0.5$\lambda$\\
			\hline
			Beam azimuth angle $\phi_b$ & $90,~\forall b \in \{1,\ldots,8\}$ $97,~\forall b \in \{9,\ldots,12\}$\\
			\hline
			Beam elevation angle $\theta_b$ & $ -52.5+15(b-1),~\forall b \in \{1,\ldots,8\} $\\
			& $ -45+30(b-8),~\forall b \in \{9,\ldots,12\} $ \\
			\hline
			Beamforming gain model & Fitting model of \cite{abstactchannel}\\
			\hline
			Rx antenna height & \SI{1.5}{m}\\
			\hline
			Rx antenna element pattern & Isotropic \\
			\hline
			Rx antenna element gain &\SI{0}{dBi} \\
			\hline
			Propagation loss & Deterministic model of \cite{SimplifiedDeterministic} \\
		  \hline
            Fast fading model & Abstract model of \cite{abstactchannel}\\
            \hline
            Total number of UEs $N_\textrm{UE}$ & 320\\
			\hline
			Simultaneously scheduled beams per cell $K_b$ & 4\\
			\hline
			A3 handover offset $o_{c_0,c}^\textrm{A\tsub3}$ & $3$ dB\\
			\hline
			CHO \textit{Add} offset $ o_{c_0,c}^\textrm{add} $ & \SI{-3}{dB}\\
			\hline
			CHO \textit{Execute offset} $ o_{c_0,c}^\textrm{exec} $ & \SI{3} {dB}\\
            \hline
   		Handover preparation time $ T_\textrm{p} $ & \SI{50}{ms} \\
			\hline
			L1 measurement period $ \omega $ & 2 \\
			\hline
			Nr. of averaged samples $ N_\textrm{L1} $ & 4 \\
            \hline
            Time step size $\Delta t$ & \SI{10}{ms}\\
            \hline
            SSB periodicity & \SI{20}{ms}\\
            \hline
            Simulated time & \SI{300}{s}\\ 
			\hline
            SINR threshold $\gamma_{\textrm{out}}$ &  \SI{-8}{dB} \\
            \hline        
		\end{tabulary}

	\vspace{-1\baselineskip} 
	\end{table}

\begin{figure}[!t]
		\centering
        \includegraphics[width=\columnwidth]{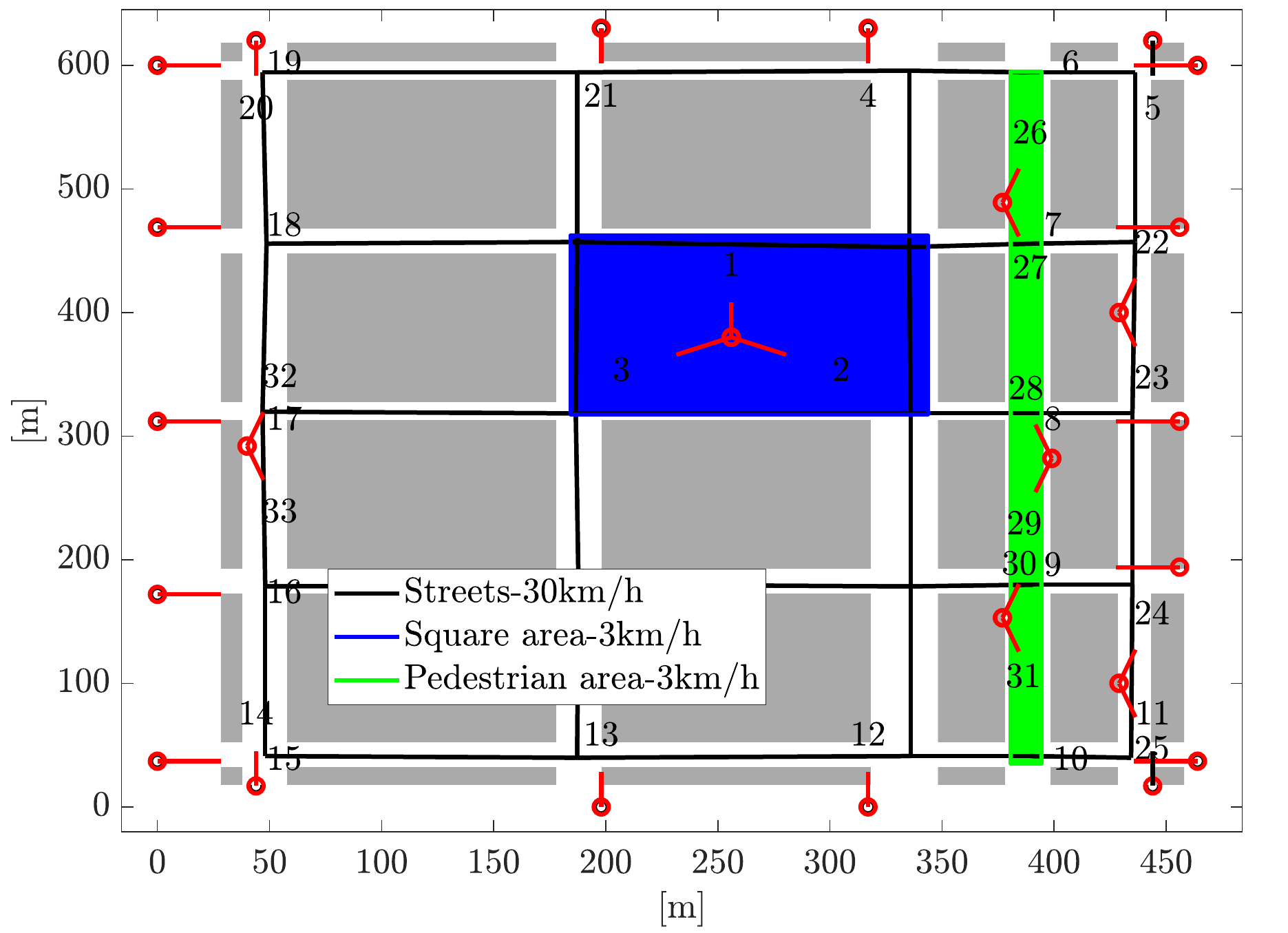}
        \vspace{-20pt}
		\caption{Madrid Grid layout is used for simulation scenario as described in METIS 2 project \cite{METIS2}. The scenario consists of buildings (grey), streets (black) with 200 users, open square (blue) with 40 users and pedestrian area (green) with 80 users.}
		\label{fig:environment}
        \vspace{-0.4\baselineskip}
	\end{figure}

	A 5G network based on the Madrid Grid layout  described in the METIS 2 project \cite{METIS2} is considered in this study. The layout is given in Fig.\,\ref{fig:environment} and consists of buildings (shown in grey), streets (show in black), an open square (shown in blue), and a pedestrian area (shown in green). There are 33 three-sector macro cells which are located on the rooftops of the buildings (shown in red). The scenario involves mixed UE traffic and users are distributed into three different categories. 200 users are moving in the streets at $30$ km/h in both directions. 40 pedestrian users are walking in the open square and 80 users are walking in the pedestrian area, both at $3$ km/h where they walk straight in a random direction and bounce when they reach the area border. The channel model used \cite{38901} takes into account shadow fading due to large obstacles and assumes a soft line-of-sight for all radio links between the cells and UEs. Fast fading is taken into consideration through the low complexity channel model for multi-beam systems proposed in \cite{abstactchannel}, which integrates the spatial and temporal characteristics of 3GPP’s geometry-based stochastic channel model \cite{38901} into Jake’s channel model. The transmitter (Tx)-side beamforming gain model is based on~\cite{abstactchannel}, where a 12-beam grid configuration is considered. Beams $ b\in \{1,\ldots,8\}$ have smaller beamwidth and higher beamforming gain to cover an area further apart from the BS. Whereas beams $ b \in \{9,\ldots,12\}$ with larger beamwidth and relatively smaller beamforming gain are defined to serve regions closer to the BS.  The number of simultaneously scheduled beams per cell is taken as $K_b=$ 4. 

 The average dowlink SINR $ \gamma_{c,b}(m) $ of a link between the UE and beam $ b $ of cell $ c $ is evaluated by the approximation given in \cite{SINRModel} for the strict resource fair scheduler, where all UEs get precisely the same amount of resources. This SINR is then used in deriving the HOF and radio link failure (RLF) models, each of which are discussed~below.
	
	\textit{Handover Failure Model}: The HOF model is used to determine the failure of a UE to hand over from its serving cell $c_0$ to its target cell $c_\textrm{T}$. As shown in Fig.\,\ref{fig:RACHflow}, for both 3GPP and the proposed RE-RACH procedure, the UE may decide to use either CBRA or CFRA preamble and attempt to access the selected prepared beam $ b_0 $ of target cell $c_\textrm{T}$  with the selected preamble. For successful random access, it is required that the SINR $ \gamma_{c_\textrm{T},b_0}(m) $ of the target cell remains above the threshold $ \gamma_\textrm{out} $, during the RACH procedure. A HOF timer $ T_\textrm{HOF}= \SI{500}{ms}$ is started when the UE starts the random access and sends the RACH preamble. The RACH procedure in Fig.\,\ref{fig:RACHflow} is repeated until a successful RACH attempt is achieved or $ T_\textrm{HOF} $ expires. A UE may succeed in accessing the target cell only if $ \gamma_{c_\textrm{T},b_0}(m) $ exceeds the threshold $ \gamma_{\textrm{out}} $. In case the timer $ T_\textrm
	{HOF} $ expires and the UE fails to access the target cell, i.e., $ \gamma_{c_\textrm{T},b} < \gamma_\textrm{out} $, a HOF is declared. The UE then performs connection re-establishment, which results in additional signaling and handover interruption~time \cite{38331}.
	
	\textit{Radio Link Failure Model}: The RLF model is used to determine failure of a UE while in its serving cell $c_0$. The UE is prompted to start an RLF timer $T_\textrm{RLF}= \SI{600}{ms}$ when the SINR  $\gamma_{c_0,b}(m)$ of the serving cell $ c_0 $ falls below $\gamma_\textrm{out}$. An RLF is declared if $T_\textrm{RLF}$ expires. During the time, the UE may recover before declaring an RLF if the SINR $ \gamma_{c_0,b} $ exceeds a second SINR threshold defined as $\gamma_\textrm{in}$ = \SI{-6}{dB}, where $\gamma_\textrm{in} > \gamma_\textrm{out}$ . A more detailed explanation of the procedure can be found in~\cite{38331}.
	
	\section{Performance Evaluation}
	\label{sec:PerformanceEval}
	In this section, the mobility performance of the proposed RE-RACH procedure is compared with 3GPP RACH for both BHO and CHO. The performance analysis is then extended for the BELL approach. The key performance indicators (KPIs) used for comparison are explained below. 
	
	\subsection{Mobility KPIs}
	\label{sec:KPIs}
	
	\subsubsection{CBRA Ratio ($R_\textrm{CBRA}$)} The ratio of the total number of CBRA events to the sum of the total number of CBRA and CFRA events in the network, denoted as $N_\textrm{CBRA}$ and $N_\textrm{CFRA}$, respectively. The ratio is thus defined as
	\begin{equation}\label{CBRA rate}
	R_\textrm{CBRA}[\%] = \frac{N_\textrm{CBRA}}{N_\textrm{CBRA}+N_\textrm{CFRA}}\cdot 100\%.
	\end{equation}
	\subsubsection{$ N_\textrm{HOF}$} The total number of HOFs that are declared in the network, following the HOF model discussed in Section~\ref{sec:SimRes}. 
	
	\subsubsection{$ N_\textrm{RLF}$} The total number of RLFs that are declared in the network, following the RLF model discussed in Section~\ref{sec:SimRes}. 
	
	Both $ N_\textrm{HOF} $ and $ N_\textrm{RLF} $ are normalized to the total number of UEs $N_\textrm{UE}$ in the network per minute and expressed as UE/min.

	\subsection{Simulation Results}
	\label{sec:SimRes}
	
	The mobility performance of the 3GPP and RE-RACH procedures is investigated for both CHO and BHO. To this end, the impact of different beam access thresholds $ \xi_\textrm{access} $ and numbers of prepared beams $ N_\textrm{B}$ are analyzed in terms of the mobility KPIs discussed~above. 
 
	


	\subsubsection{3GPP and RE-RACH Performance with CHO}
	\label{sec:3GPP-RE-RACH-CHO}
\begin{figure}[!t]
         \centering 
        \subfigure[Ratio $R_\textrm{CBRA}$ of CBRA resource usage.]
        {
             \includegraphics[width=\columnwidth]{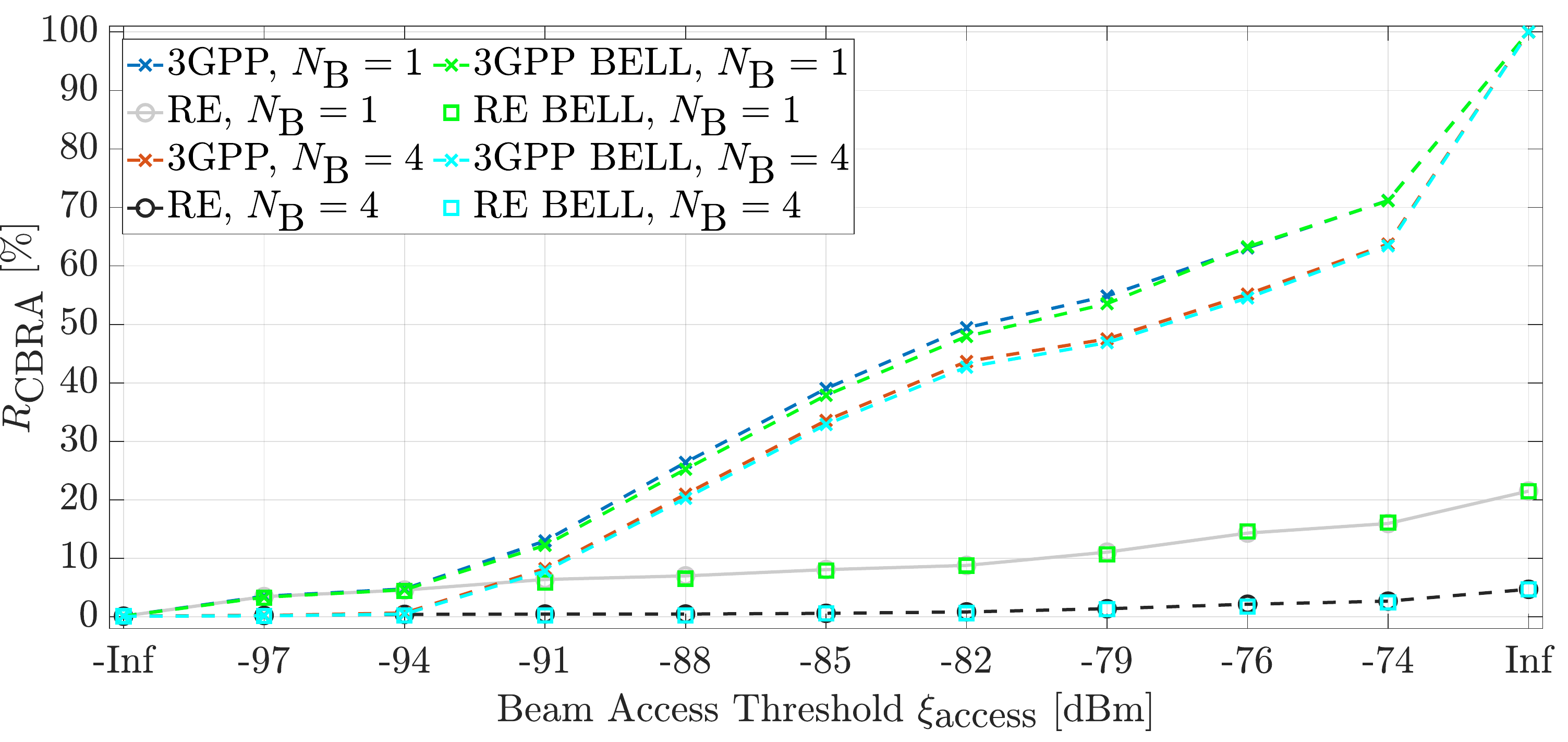}
            \label{fig:CHO_RACH}
            \vspace{-1.5\baselineskip}
        }

        \subfigure[Normalized HOFs.]
        {
            \includegraphics[width=\columnwidth]{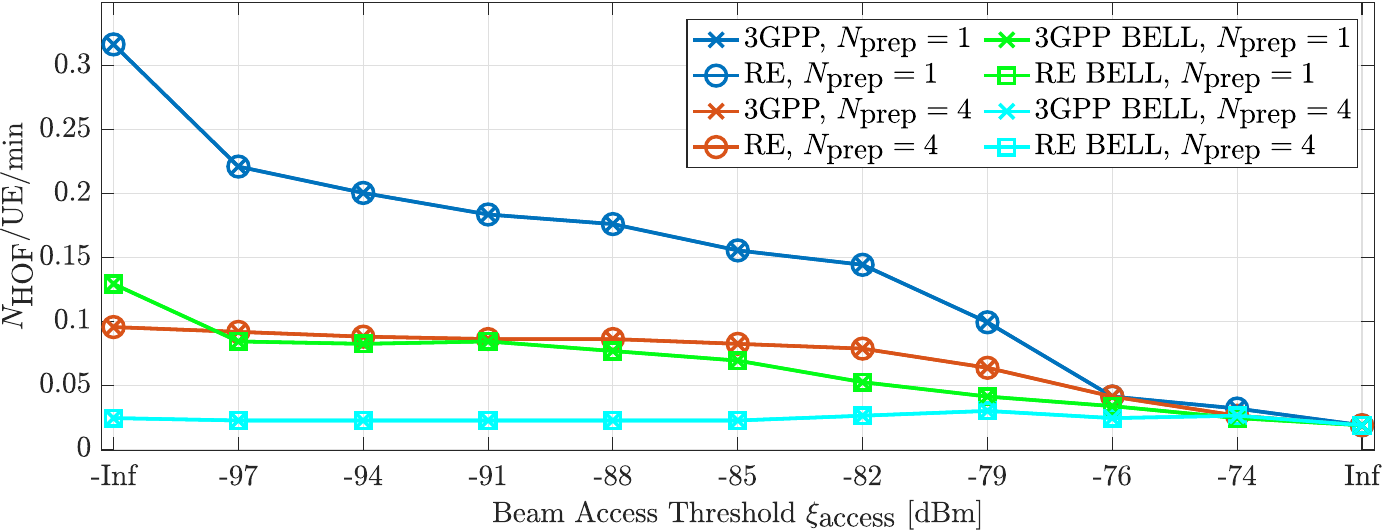}
            \label{fig:CHO_HOF}
            \vspace{-0.125\baselineskip}
        }
        
        \vspace{-6pt}
        \caption{The mobility performance of the 3GPP and RE-RACH procedures for CHO, shown as a function of the beam access threshold $\xi_\textrm{access}$. The mobility performance is shown in terms of (a) the ratio $R_\textrm{CBRA}$ and (b) HOFs for different numbers of prepared beams for random access $N_\textrm{B}$. The performance of the BELL approach is also shown for both the RACH procedures.}
        \label{fig:CHO} 
\end{figure}

    It can be seen in Fig.\,\ref{fig:CHO_RACH} that for $ \xi_\textrm{access}$ = $-\infty$ the ratio $ R_\textrm{CBRA}$ equals zero for all the six cases in consideration, i.e., the UE uses only CFRA preambles during RACH. This is because the UE always selects a prepared beam from the $ B^\textrm{prep}_{c_\textrm{T}} $ to perform random- access d towards the target cell $ c_\textrm{T} $, as given in (\ref{eq:SelectBestPrep}). As per (\ref{eq:SelectStrongBeam}), for higher $ \xi_\textrm{access} $ values the UE tends to select prepared beams less frequently, which results in more likely use of CBRA preambles for random access with the 3GPP approach. It is also clearly observed  in Fig.\,\ref{fig:CHO_RACH} that the ratio of CBRA resource usage is much smaller for the RE-RACH procedure for higher values of $ \xi_\textrm{access} $. This is because the UE now has the option of using CFRA resources if none of the prepared beams have beam measurements above the threshold $ \xi_\textrm{access} $. On the other hand, it can be seen in Fig.\,\ref{fig:CHO_HOF} that the beam access threshold $ \xi_\textrm{access} = -\infty$ has the worst HOF performance for all the six cases under consideration. This is because the RSRP of the prepared beam changes over time and the prepared beam does not always remain a good candidate during the time that has elapsed between the handover preparation and execution phases. Ultimately, the SINR $ \gamma_{c,b_0}(m) $ of the accessed beam $ b_0 $, where $ m\geq m_1 $, falls below $ \gamma_\textrm{out} $ which leads to a HOF. This is more visible for $ N_\textrm{B}=1 $ since the UE does not have any other options for selecting another prepared beam. Increasing $ N_\textrm{B} $ from $ 1 $ to $ 4 $ reduces the access failure $N_\textrm{HO}$ to almost one-third of its value because it increases the chance of the strongest beam being selected by the UE.

	For increasing values of access threshold $ \xi_\textrm{access} $, the RACH beam selection procedure prioritizes the L1 RSRP beam measurements $ P_{c,b}^\textrm{L1}(m) $ as given in (\ref{eq:SelectStrongBeam}) and the UE becomes less persistent on selecting one of the prepared beams and instead simply selects the beam with the strongest L1 RSRP beam measurement. As a consequence, beams with higher $ P_{c,b}^\textrm{L1}(m) $ are selected to be accessed, which yields a higher SINR $ \gamma_{c_0,b}(m)$ and consequently fewer HOFs. This can be seen in Fig.\,\ref{fig:CHO_HOF}. It can also be visualized in Fig.\,\ref{fig:CHO_HOF} that the same HOF performance is observed for both the 3GPP and the proposed RE-RACH procedures for the respective number of prepared beams $N_\textrm{B}$ cases. This is due to the fact that the RE-RACH procedure only focuses on increasing CFRA resource usage whereas the HOF performance is dependent on the selected access beam $b_0$ that is taken into account in the target cell SINR given in the HOF model Section~\ref{sec:SimScenario}.
	

	\subsubsection{BELL Performance with CHO}
	It can be seen in Fig.\,\ref{fig:CHO_HOF} that the total number of normalized HOFs of both the 3GPP and RE-RACH procedures without the BELL approach is approximately $ 0.3 $/UE/min for $ \xi_\textrm{access}=-\infty $ and $ N_\textrm{B}=1 $ (shown in blue). Applying the BELL approach on both the RACH procedures prevents around $ 60\% $ the HOFs and reduces them to around $ 0.13 $/UE/min (shown in green). For beam access threshold values $ \xi_\textrm{access} > -\infty $, the same trend is seen for both $N_\textrm{B}=1 $ and $ N_\textrm{B}=4 $ and the results converge as $ \xi_\textrm{access}$ approaches $\infty $. This shows that the BELL approach not only improves the HOF performance but also requires less resources to be reserved since essentially the same performance can be improved by preparing less beams $ N_\textrm{B} $ for higher $ \xi_\textrm{access}$ values. Furthermore, it is seen that for $ N_\textrm{B}=4 $ the HOF performance of the BELL approach (shown in cyan) remains fairly constant for all $ \xi_\textrm{access} $ values and shows the same performance as that seen at $ \xi_\textrm{access} = \infty$, where always the strongest beam is selected as the access beam, no matter which beam was prepared. Hence, the efficacy of the scheme can be seen here. It is to be noted here that the BELL approach does not react on the other classes that are defined for HOFs in Fig.\,\ref{fig:decisiontree}, such as coverage holes, early execution, and wrong cell preparation but only focuses on addressing HOFs caused by wrong beam preparation. This is the cause of residual failures seen at $ \xi_\textrm{access} =\infty$ for the BELL approach.

    It is also seen in Fig.\,\ref{fig:CHO_RACH} that ratio $ R_\textrm{CBRA} $ is slightly reduced when the BELL approach is employed. This is because the BELL approach designates the prepared beam based on a rewarding algorithm that leverages the beams with higher L1 RSRP beam measurements at the time of access $ m>m_1 $. Therefore, the chance of preparing a beam is indirectly increased, where the prepared beams are more likely to be CFRA on account of their higher RSRP, as given in (\ref{eq:SelectBestPrep}).
	
	Lastly, it can be concluded that the BELL approach with $ N_\textrm{B}=4 $  provides an optimum operation point at $ \xi_\textrm{access} =-\infty$ for CHO. This is because it gives full CFRA resource utilization ($ R_\textrm{CBRA}=0\% $) for minimum achievable HOF, where the HOFs are comparable to the minimum convergence value seen at  $\xi_\textrm{access} =-\infty$ for all the six cases under consideration.

	\subsubsection{RACH Performance Comparison between BHO and CHO}
\begin{figure}[!t]
    		  \centering
     	 \includegraphics[width=\columnwidth]      
             {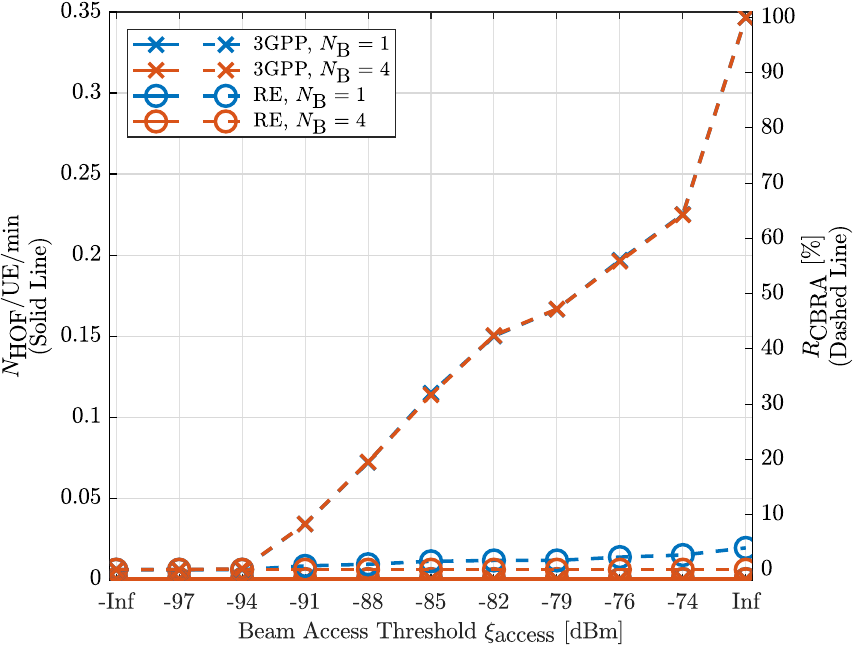}            
               \vspace{-14pt}
    		  \caption{The mobility performance of the 3GPP and RE-RACH procedures for BHO, shown as a function of the beam access threshold $\xi_\textrm{access}$. The mobility performance is shown in terms of HOFs (solid line) and ratio $R_\textrm{CBRA}$ (dashed line), for different numbers of prepared beams for random access $N_\textrm{B}$.}
    		  \vspace{-1.2\baselineskip}
             \label{fig:BHO} 
    	\end{figure}

	It can be observed in Fig.\,\ref{fig:BHO} that HOFs (shown in solid line) are not observed in BHO for any number $ N_\textrm{B} $ of prepared beams and beam access threshold $ \xi_\textrm{access} $. As explained in Section~\ref{subsec:AccessPreamble}, this is because the time $ T_\textrm{p} $ that elapses between handover preparation and execution in BHO is shorter than that of CHO ($T_\textrm{p} + T_\textrm{f} $) and during this time the measurements of the prepared beams do not undergo large changes. Consequently, the UEs perform access to a beam $ b_0 $ that yields a sufficient SINR $ \gamma_{c_\textrm{T},b_0}(m) $ at target cell $ c_\textrm{T} $. The BELL approach cannot improve the HOF performance of BHO since it focuses on reducing the HOFs, which to begin with are not observed~here.

	It can also be seen  in Fig.\,\ref{fig:BHO} that the CBRA ratio (shown in dashed line)  of the 3GPP procedure for increasing $ \xi_\textrm{access} $ values sees a similar trend as seen for CHO in Fig.\,\ref{fig:CHO}. This is again due to the fact that the 3GPP RACH procedure does not consider the prepared beams in case the L1 measurements are below the access threshold~$ \xi_\textrm{access}$. Fig.\,\ref{fig:BHO} also shows that the CBRA ratio of the proposed RE-RACH procedure is close to 0\% and only slightly increases for higher $ \xi_\textrm{access} $ because the measurements of the beams do not change much in the time interval between preparation and execution phases, which is much shorter than that of the CHO case.

	\subsubsection{Total Number of Mobility Failures}
\begin{figure}[!b]
		\centering
        \includegraphics[width=0.9\columnwidth]{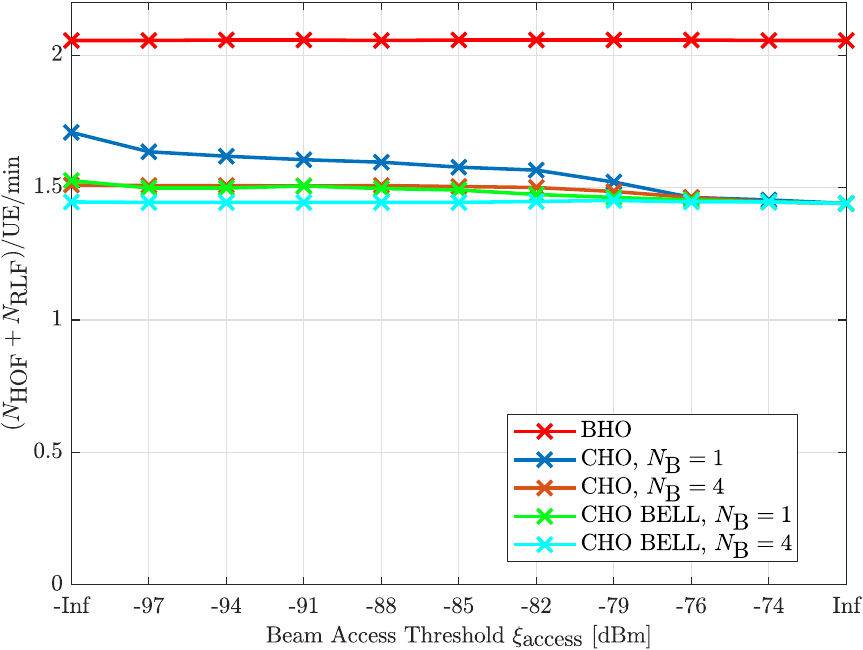}
        \vspace{-8pt}
		\caption{A comparison of the mobility performance of CHO with BHO in terms of the total mumber of mobility failures, shown as a function of beam access threshold $ \xi_\textrm{access} $ with different numbers of prepared beams for random access $N_\textrm{B} $.}
		\label{fig:RLF}
	\end{figure}

	Fig.\,\ref{fig:RLF} shows a comparison between BHO and CHO for the total number of failures, taken as a sum of $ N_\textrm{HOF}$ and $N_\textrm{RLF}$. Since it has been shown in Figs.\,\ref{fig:CHO} and \ref{fig:BHO} that the HOFs are independent of the RACH procedure, the results in Fig.\,\ref{fig:RLF} do not differentiate between them. Furthermore, the total number of failures for BHO results in the same performance for both $ N_\textrm{B}=1 $ and $ N_\textrm{B}=4 $ and for the BELL approach, and therefore a differentiation has not been~made.
	
	Although Fig.\,\ref{fig:BHO} shows that the mobility performance of BHO in terms of  HOF is better than that of CHO as seen in  Fig.\,\ref{fig:CHO_HOF}, Fig.\,\ref{fig:RLF} illustrates that the overall mobility failure performance of BHO is improved by the conditional execution mechanism that is introduced by CHO. When BHO (in red) is compared with CHO BELL, $N_\textrm{B}=4$ (in cyan) at $\xi_\textrm{access} =\infty$, a relative reduction of $ 30\% $ is seen (from $ 2.06 $ to $1.45$/UE/min). Furthermore, one can also state that the mobility failures observed in the mobility scenario are dominated by RLFs, which is improved by CHO. The enhancements of a larger number of prepared beams for RACH $ N_\textrm{B} $ and BELL approach on the overall performance of CHO can also be seen.

	\section{Conclusion}
	\label{sec:Concl}
	
	In this article, conditional handover of \textit{3GPP Release 16} is analyzed for a 5G beamformed network. Baseline and conditional handover procedures have been reviewed along with L1 and L3 UE measurements that are imperative for mobility. In addition, the 3GPP random access procedure is revisited and a new random access procedure is proposed that aims to increase contention-free random access (CFRA) and in turn reduce the signaling overhead and latency during handover. Besides, a decision tree-based supervised learning method is proposed that reduces the handover failures (HOFs) caused by the beam preparation phase of the RACH procedure. The results show that the optimum operation point is achieved with the proposed learning algorithm. Furthermore, the mobility performance of conditional handover is compared with baseline handover. Simulation results have shown that the number of fall-backs to CBRA is reduced significantly when the proposed random access procedure is used. Moreover, the results have shown that the baseline handover procedure causes fewer handover failures than  conditional handover. However, the total number of failures for conditional handover is less than that of baseline handover due to the decoupling of  handover preparation and execution phases. Based on these findings, future studies may be conducted to investigate the effect of the proposed schemes with enhanced conditional handover techniques \cite{fastchopaper}.

%
%
  
	\bibliographystyle{IEEEtran}
    \vspace{-0.1\baselineskip}
	\bibliography{references}

\begin{thebibliography}{10}
\providecommand{\url}[1]{#1}
\csname url@samestyle\endcsname
\providecommand{\newblock}{\relax}
\providecommand{\bibinfo}[2]{#2}
\providecommand{\BIBentrySTDinterwordspacing}{\spaceskip=0pt\relax}
\providecommand{\BIBentryALTinterwordstretchfactor}{4}
\providecommand{\BIBentryALTinterwordspacing}{\spaceskip=\fontdimen2\font plus
\BIBentryALTinterwordstretchfactor\fontdimen3\font minus
  \fontdimen4\font\relax}
\providecommand{\BIBforeignlanguage}[2]{{%
\expandafter\ifx\csname l@#1\endcsname\relax
\typeout{** WARNING: IEEEtran.bst: No hyphenation pattern has been}%
\typeout{** loaded for the language `#1'. Using the pattern for}%
\typeout{** the default language instead.}%
\else
\language=\csname l@#1\endcsname
\fi
#2}}
\providecommand{\BIBdecl}{\relax}
\BIBdecl
\renewcommand{\BIBentryALTinterwordstretchfactor}{4}

\bibitem{Throughput}
Cisco, ``{Cisco Annual Internet Report (2018–2023)},'' Tech. Rep., Mar 2020,
  {White Paper C11-741490-01}.

\bibitem{38331}
3GPP, ``{NR}; radio resource control protocol specification,'' {3rd Generation
  Partnership Project (3GPP)}, Tech. Rep. 38.331, Jun. 2022, {V16.9.0}.

\bibitem{HOsurvey}
M.~{Tayyab}, X.~{Gelabert}, and R.~{Jäntti}, ``A survey on handover
  management: From lte to nr,'' \emph{IEEE Access}, vol.~7, pp.
  118\,907--118\,930, 2019.

\bibitem{38300}
3GPP, ``{NR; NR and NG-RAN} overall description stage-2,'' {3rd Generation
  Partnership Project (3GPP)}, Tech. Rep. 38.300, Sep. 2021, {V16.7.0}.

\bibitem{CHOTechRep}
3GPP, ``Conditional handover – basic aspects and feasibility in {Rel-15},''
  {3rd Generation Partnership Project (3GPP)}, Tech. Rep. TSG-RAN WG2 NR Adhoc
  2, Jun 2017, {R2-1706489}.

\bibitem{RAlinktoSingalingOverheadandLatency1}
E.~Peralta, T.~Levanen, F.~Frederiksen, and M.~Valkama, ``Two-step random
  access in 5g new radio: Channel structure design and performance,'' in
  \emph{IEEE VTC-Spring}, 2021, pp. 1--7.

\bibitem{RAlinktoSingalingOverheadandLatency2}
Z.~Li, L.~Tian, Y.~Yin, and W.~Cao, ``On contention-based 2-step random access
  procedure,'' in \emph{2020 International Conference on Wireless
  Communications and Signal Processing (WCSP)}, 2020, pp. 771--776.

\bibitem{JedrzejRACHpaper}
J.~Stanzcak, U.~Karabulut, and A.~Awada, ``Conditional handover modelling for
  increased contention free resource use in 5g-advanced,'' \emph{in IEEE
  PIMRC}, pp. 1--6, 2023, accepted for publication, available:
  https://arxiv.org/abs/2307.14870.

\bibitem{MarvinHOFpaper}
M.~Manalastas, M.~U.~B. Farooq, S.~M.~A. Zaidi, A.~Abu-Dayya, and A.~Imran, ``A
  data-driven framework for inter-frequency handover failure prediction and
  mitigation,'' \emph{IEEE Transactions on Vehicular Technology}, vol.~71,
  no.~6, pp. 6158--6172, 2022.

\bibitem{SupervisedLearning}
T.~Hastie, R.~Tibshirani, and J.~Friedman, \emph{The Elements of Statistical
  Learning - Data Mining, Inference, and Prediction}.

\bibitem{METIS2}
A.~Weber \emph{et~al.}, ``Performance evaluation framework,'' {Mobile and
  wireless communications Enablers for the Twentytwenty Information
  Society-II}, Tech. Rep., 2016, {Deliverable D2.1}.

\bibitem{38901}
3GPP, ``Study on channel model for frequencies from 0.5 to 100 {GHz},'' {3rd
  Generation Partnership Project (3GPP)}, Tech. Rep. 38.901, Dec. 2019,
  {V16.1.0}.

\bibitem{abstactchannel}
U.~Karabulut, A.~Awada, I.~Viering, A.~N. Barreto, and G.~P. Fettweis, ``Low
  complexity channel model for mobility investigations in 5g networks,'' in
  \emph{IEEE WCNC}, 2020, pp. 1--8.

\bibitem{SimplifiedDeterministic}
A.~{Awada}, A.~{Lobinger}, A.~{Enqvist}, A.~{Talukdar}, and I.~{Viering}, ``A
  simplified deterministic channel model for user mobility investigations in 5g
  networks,'' in \emph{IEEE ICC}, 2017, pp. 1--7.

\bibitem{SINRModel}
A.~{Ali} \emph{et~al.}, ``System model for average downlink sinr in 5g
  multi-beam networks,'' in \emph{IEEE PIMRC}, 2019, pp. 1--6.

\bibitem{fastchopaper}
S.~Bin~Iqbal \emph{et~al.}, ``On the modeling and analysis of fast conditional
  handover for 5g-advanced,'' in \emph{IEE PIMRC}, 2022, pp. 595--601.

\end{thebibliography}

\end{document}